\documentclass[twocolumn]{revtex4}

\usepackage{amsfonts}
\usepackage{amsmath}
\usepackage{amssymb}

\usepackage{ragged2e}
\usepackage{graphicx}
\usepackage[usenames,dvipsnames]{color}
\definecolor{darkblue}{RGB}{0,0,196}

\begin{document}

\title{Extraction of freezeout parameters and their dependence on collision energy and collision cross-section \vspace{0.5cm}}

\author {{Muhammad Waqas$^{1,}$\footnote{waqas\_phy313@yahoo.com, waqas\_phy313@ucas.ac.cn},
Guang Xiong Peng$^{1,}$\footnote{Correspondence: gxpeng@ucas.ac.cn},
Muhammad Ajaz$^{2,}$, 
Abd Al Karim Haj Ismail$^{3,4}$,\footnote{Correspondence: a.hajismail@ajman.ac.ae},
Pei-Pin Yang$^{5}$, 
Zafar Wazir$^{6}$} 
\vspace{0.25cm}}

\affiliation{$^1$School of Nuclear Science and Technology, University of Chinese Academy of Sciences, Beijing 100049, Peoples Republic of china,
\\
$^2$Department of Physics, Abdul Wali Khan University Mardan, 23200 Mardan, Pakistan,
\\
$^3$Department of Mathematics and Science, Ajman University, PO Box 346, UAE,
\\
$^4$Nonlinear Dynamics Research Center (NDRC), Ajman University, PO Box 346, UAE,
\\
$^5$Institute of Theoretical Physics and Department of Physics, and State Key Laboratory of Quantum Optics and
Quantum Optics Devices, Shanxi University, Taiyuan, Shanxi 030006, Peoples Republic of China,
\\
$^6$Department of physics, Ghazi University, Dera Ghazi Khan, Pakistan}

\begin{abstract}

\vspace{0.5cm}

\noindent {\bf Abstract:} We used the Blast wave model with Boltzmann Gibbs
statistics and analyzed the experimental data of transverse momentum spectra
($p_T$) measured by NA61/SHINE and NA 49 Collaborations in inelastic (INEL)
proton-proton, and the most central Beryllium-Beryllium (Be-Be),
Argon-Scandium (Ar-Sc) and Lead-Lead (Pb-Pb) collisions. The model results fit the
experimental data of NA61/SHINE and NA 49 Collaborations very well. We extracted
kinetic freezeout temperature, transverse flow velocity and kinetic freezeout volume
directly from the spectra. We also calculated mean transverse momentum and initial
temperature from the fit function. It is observed that the kinetic freezeout temperature increases with increasing
the collision energy as well as collision cross-section (size of the colliding system).
Furthermore, the transverse flow remains unchanged with increasing the collision energy, while it
changes randomly with the collision cross-section. Similarly, with the increase in collision
energy or the collision cross-section, the freeze-out volume and the average $p_T$ increase.
The initial temperature is also observed to be an increasing function of the collision cross-section.
\\

\end{abstract}

\maketitle

\section{Introduction}
A new state of matter, namely Quark-Gluon Plasma (QGP), is formed under the extreme conditions of temperature
and energy densities. This matter is formed in the early stages of the collisions and have a very short
lifetime  of almost 7-10 fm/c, after which it changes quickly to a system of hadron gas. Due to multi-
partonic interactions in the collision system, the information about the initial condition gets lost
and we can get the final state behavior of such systems from measure of numbers and identity of produced
particles along with their energy and transverse momentum spectra.

During heavy ion collisions, the constituents of hot and dense matter interact elastically or inelastically
with each other and evolves into a new state of free matter. This phenomenon of particles decoupling is
called freezeout. There are two kinds of freezeout. When the colliding medium reaches a stage of chemical
equilibrium and then cools down by expansion. The inelastic collisions stop due to the expansion of the system,
and the mean free path for the interactions becomes comparable to the size of the system. In addition, the abundances
of different particle species get fixed. This is referred as chemical freezeout. The chemical freezeout stage
is followed by another stage, although the relative fractions of the particles are fixed, they continue
to interact with each other until the final state interactions between the particles are no longer effective. This
is known as kinetic freezeout. At the stage of kinetic freezeout, all the interactions stop and the transverse
momentum spectra of the particles do not change. Therefore, the transverse momentum spectra of the particles
are very important due to the reason that it contains the necessary information about the final state particles,
including the kinetic freezeout temperature ($T_0$), transverse flow velocity ($\beta_T$), kinetic freezeout
volume ($V$) and time of flight of the particles.

The above discussed freezeout stages correspond to the chemical freezeout temperature ($T_{ch}$) and kinetic
freezeout temperature ($T_0$) respectively. In addition to the two kind of temperatures, the initial temperature ($T_i$) is also
important because of its determining effect to the evolution system.

Large collision systems are expected to form QGP matter, however small systems (especially with low multiplicity)
are expected to create it due to the reason that they have a small volume of violent collision regions. From the similar
multiplicity at the energy range up to 200 GeV, small systems are more similar to peripheral AA collisions, however, they
are not similar to central AA collisions [1]. At low energies from a few GeV to 10 GeV, the situation is different
due to the dominance of baryons [2].

In the present work, we will use the Blast Wave model with Boltzmann Gibbs statistics [3, 4, 5] to analyze the transverse momentum
spectra of the pion in different collisions system at different energies and will extract the kinetic freezeout temperature,
transverse flow velocity and kinetic freezeout volume. We will show the dependence of the above parameters on collision
energy and collision cross-section. The dependence of these parameters are in contradiction in different literatures [3, 5, 6, 7, 8, 9, 10, 11, 12, 13, 14, 15, 16, 17, 18, 19].
In fact, it is very useful to do more studies on these topics and to finalize some corroborative conclusions. In addition, we will
also extract the initial temperature ($T_i$) and average transverse momentum ($<p_T>$) from the fit function to show their
dependence on collision energy and collision cross-section.

Before going to the next section, we would like to point out that we choose the Blast Wave model with Boltzmann Gibbs
statistics in the present work because it is very close to ideal gas model, and the reason behind the selection of pion
is that the temperature obtained from its spectra is the closest to the source temperature.

\section{Formalism and method}

In high energy collisions, the $p_T$ spectra of the produced particles is very complex. Although the
function can be of various forms, it is not enough to use a single probability density function to describe the
$p_T$ spectra (especially when the maximum $p_T$ is up to 100 GeV/c) [20]. There are various $p_T$ regions [21] according
to the model analysis that are described in our previous work [22]. The soft excitation and hard scattering
are the two main processes for particle production. The soft excitation process results in the production of most light
flavor particles whose $p_T$ range is narrow (less than 2--3 GeV/c), while hard scattering process exists in a wide
$p_T$ range ($p_T$ $>$3 GeV/c), in addition, some light flavor particles are also produced in this process. In some cases of not too
high collision energies, the hard scattering process can be underestimated and the soft excitation plays the main role in
the particle production. Similar or different probability density functions describe the soft and hard scattering processes.
In general, due to the small fraction of the hard process in narrow $p_T$ range, the hard scattering process does not contribute
in temperature and flow velocity. For soft excitation process, we have various choices of formalisms which include but are
not limited to the standard distribution [23], Tsallis statistics [24--26], Erlang distribution [27--29], Schwinger mechanism [30--33],
the Blast Wave model with Boltzmann Gibbs statistics [4 ,34], Blast Wave model with Tsallis statistics [35--37] and Hagedorn
thermal model distribution [38].

In the present work, we used the Blast Wave model with Boltzmann Gibbs statitics (BGBW), which assumes that the particles are
locally thermalized at the thermal/kinetic freezeout temperature, and are moving with a common transverse collective flow
velocity [4, 39]. Let us assume a thermal  source, which is radially boosted, has kinetic freezeout temperature ($T_0$) and a
transverse radial flow velocity ($\beta_T$), the transverse momentum ($p_T$) spectra distribution of the particles is

\begin{align}
f(p_T)=&\frac{1}{N}\frac{dN}{dp_T} = \frac{1}{N}\frac{gV}{(2\pi)^2} p_T m_T \int_0^R rdr \nonumber\\
& \times I_0 \bigg[\frac{p_T \sinh(\rho)}{T} \bigg] K_1
\bigg[\frac{m_T \cosh(\rho)}{T} \bigg],
\end{align}
where $m_T$ ($m_T=\sqrt{p_T^2+m_0^2}$) represents the transverse mass of the particles, $g$ is the spin degeneracy factor of the particle which is 1 for pion, $I_0$ and $K_1$ are the modified Bessel function of the first and second kind respectively. $\rho= \tanh^{-1} [\beta(r)]$ is the radial flow velocity profile, (r/R) represents
the relative radial position in thermal source. The average $\beta(r)$ can be obtained from $<\beta_T>$=$2\beta_S$/($n_0$+2), where $n_0$ is the self similar flow velocity profile and its value can be 1 [13] or 2 [39] or it may also be considered a free parameter [40]. In some cases, it is possible that BGBW does not fit the whole $p_T$ region, then we use two-component of the model, which is not discussed in the present work. However the whole methodology of the two component model is discussed in detail in our previous work [14, 22, 41].

In the fit process, the extracted parameters has usually a correlation, such as $T_0$ in some cases is larger and  a smaller $\beta_T$ can lead to a similar result if a smaller $T_0$ and larger $\beta_T$ are used.  This is due to the influence of $p_T$ range, and also $n_0$ if taken as free parameter. To reduce the effect of such correlation, we need to analyze the mean $p_T$ ($<p_T>$) and the root-mean square $p_T$ over $\sqrt{2}$ ($\sqrt{<p^2_T>/2}$). We can calculate $<p_T>$ and $\sqrt{<p^2_T>/2}$ from the fit function over a given $p_T$ range, where $\sqrt{<p^2_T>/2}$ is initial temperature of interacting system according to string percolation
model [42--44].

\section{Results and discussions}
The transverse momentum ($p_T$) spectra of $\pi^-$ mseons produced in inelastic (INEL) proton-proton (pp) , Beryllium-Beryllium (Be-Be), Argon-Scandium (Ar-Sc) and Lead-Lead (Pb-Pb) collisions at different energies are represented in fig. 1. Panel (a)-(d) show the $p_T$ spectra of pion in inelastic (INEL) p-p collision, and the most central Be-Be, Ar-Sc and Pb-Pb collisions respectively at $|y|=0.1$ rapidity. The symbols represent the experimental data of NA61/SHINE [45, 46, 47] and NA49 Collaboration [48, 49] measured at the SPS CERN, while the curve represents the result of our fit by the Blast Wave model with Boltzmann Gibbs statistics. On can see that Eq. (1) approximately describes the experimental data well {\bf and the related parameters are extracted from the fit of Eq. (1) to the data by using the least square method}. The data in fig. 1(a), 1(b) and 1(c) are taken from ref. [45], [46] and [47] respectively, while the data in panel (d) is taken from ref. [48, 49]. In order to see the fit results clearly, the spectra of pion in pp collisions at 40, 80 and 158 A GeV/c in panel (a) are scaled by 4, 14 and 40, respectively. In panel (b), at 30, 40 and 150 A GeV/c, the spectra are scaled by 4, 2 and 1/8, respectively. While in panel (c), at 40, 80 and 160 A GeV/c, the spectra are scaled by 20, 8 and 4, respectively.

The lower layer in each panel represents the corresponding ratio of data/fit. The related values of free parameters and $\chi^2$ and degrees of freedom (dof) are presented in table 1. One can see that Eq. (1) provides an approximately well fit to the data in all collisions at all energies.
\begin{figure*}
\begin{center}
\includegraphics[width=14.cm]{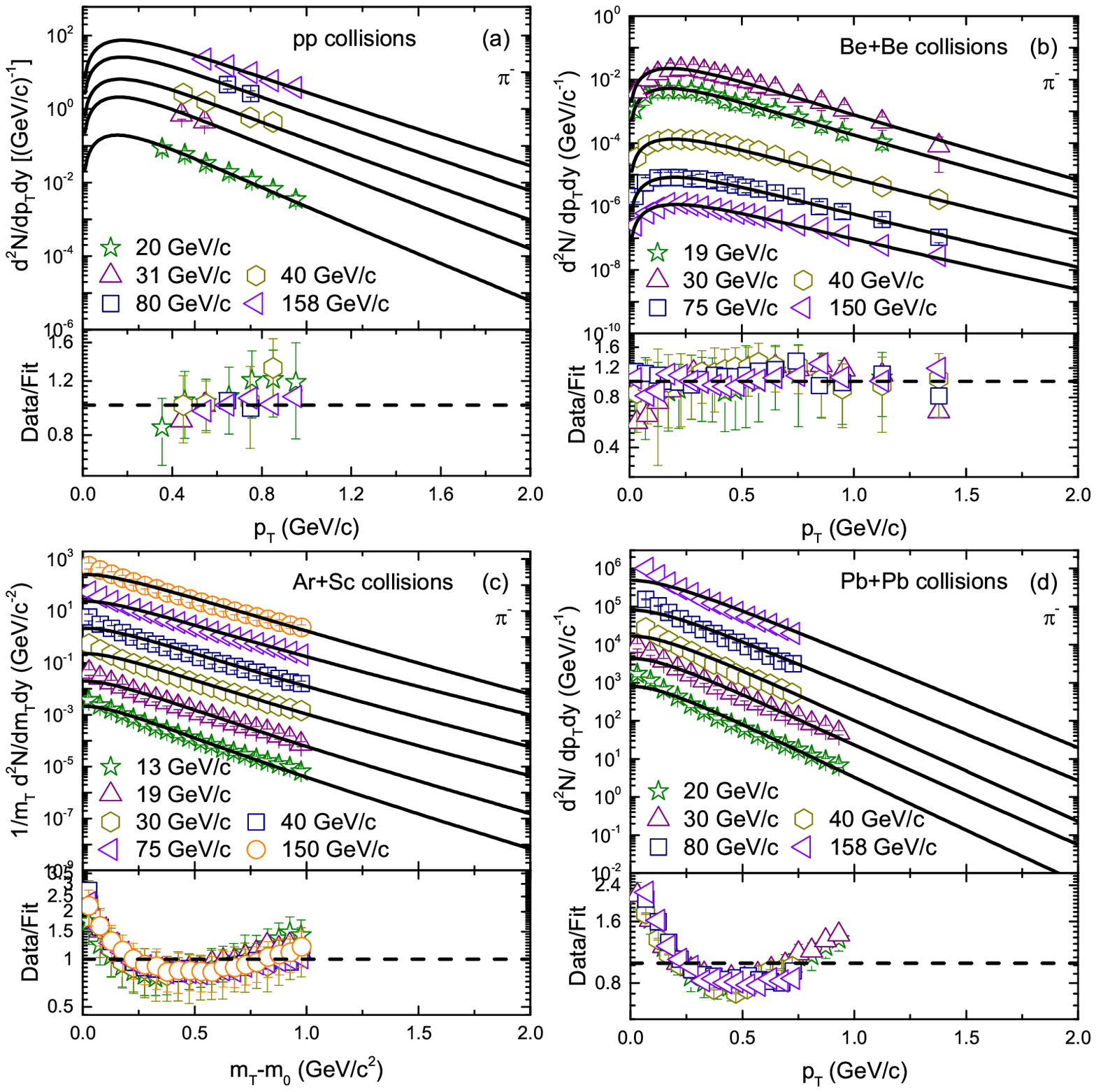}
\end{center}
Fig.1. Transverse momentum spectra of $\pi^-$ produced in pp, and the most central Be-Be, Ar-Sc and Pb-Pb collisions at different energies at $|y|=0.1$ rapidity interval. The symbols are the experimental data of NA61/SHINE [45, 46, 47] and NA49 Collaboration [48, 49] measured at the SPS CERN, and the curves are our fit by using Eq. (1). The corresponding data/fit ratios are are followed in each panel.
\end{figure*}

\begin{figure*}
\begin{center}
\includegraphics[width=14.cm]{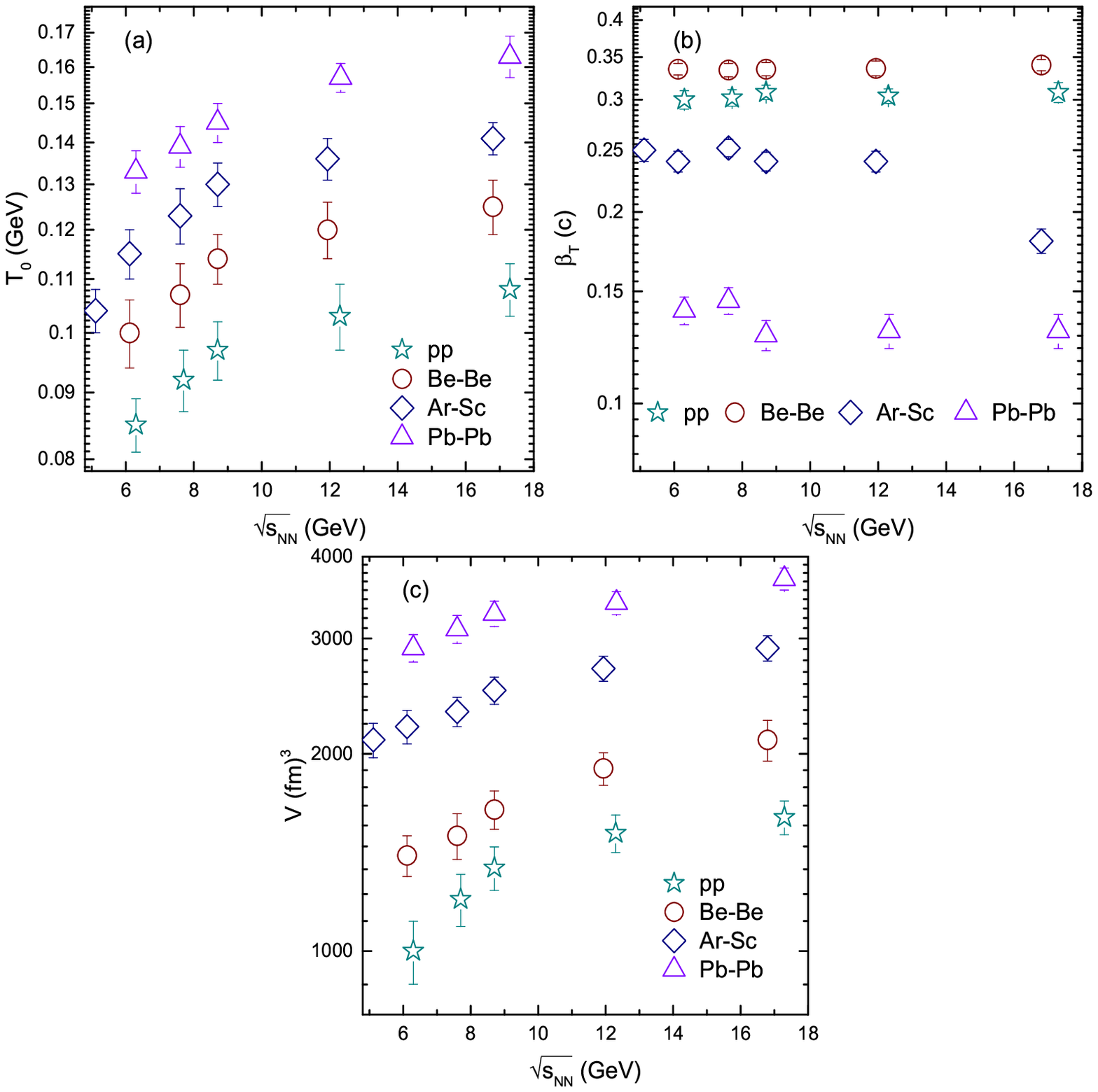}
\end{center}
Fig.2. Dependence of (a) kinetic freezeout temperature, (b) transverse flow velocity and (c) kinetic freezeout volume on collision energy and collision cross-section.
\end{figure*}

\begin{table*}
{\scriptsize Table 1. List of the parameters. (- is used in some
places instead of dof. In fact it is not the fit result. if
dof$<$0, then we put - instead of negative values.) \vspace{-.50cm}
\begin{center}
\begin{tabular}{ccccccccc}\\ \hline\hline
Collisions       & energy      & Particle     & $T_0$ (GeV)     & $\beta_T$ (c)    & $V (fm^3)$     & $N_0$      & $\chi^2$/ dof \\ \hline
Fig. 1(a)        & 20 GeV      & $\pi^-$      &$0.085\pm0.004$  & $0.300\pm0.010$  & $1000\pm110$   & $0.27\pm0.04$         & 2/3\\
  p-p            & 31 GeV      & --           &$0.092\pm0.005$  & $0.302\pm0.009$  & $1200\pm110$   &$1.4\pm0.3$            & 0.2/-\\
                 & 40 GeV      &--            &$0.097\pm0.005$  & $0.308\pm0.008$  & $1340\pm102$   & $1.9\pm0.3$           & 1/-\\
                 & 80 GeV      & --           &$0.103\pm0.006$  & $0.304\pm0.008$  & $1513\pm100$   & $2\pm0.4$             & 0.03/-\\
                 & 158 GeV     & --           &$0.108\pm0.005$  & $0.308\pm0.011$  & $1600\pm95$    & $2\pm0.4$             & 0.1/1\\
  \hline
  Fig. 1(b)      & 19 GeV      & $\pi^-$      &$0.100\pm0.006$  & $0.335\pm0.007$  & $1400\pm100$   & $0.0062\pm0.0004$     & 1/13\\
  Be-Be          & 31 GeV      & --           &$0.107\pm0.006$  & $0.334\pm0.008$  & $1500\pm120$   &$0.0063\pm0.0003$      & 5/14\\
                 & 40 GeV      &--            &$0.114\pm0.005$  & $0.335\pm0.008$  & $1645\pm110$   & $0.000075\pm0.000004$ & 3/14\\
                 & 75 GeV      & --           &$0.120\pm0.006$  & $0.336\pm0.009$  & $1900\pm108$   & $8.5E-6\pm5E-7$       & 4/14\\
                 & 150 GeV     & --           &$0.125\pm0.006$  & $0.340\pm0.007$  & $2100\pm150$   & $8.9E-6\pm4E-7$       & 1.5/14\\
  \hline
  Fig. 1(c)      & 13 GeV      & $\pi^-$      &$0.104\pm0.004$  & $0.250\pm0.010$  & $2100\pm127$   & $1.5E-4\pm4E-5$       & 13/16\\
  Ar-Sc          & 19 GeV      & --           &$0.115\pm0.005$  & $0.240\pm0.009$  & $2200\pm130$   &$0.00154\pm0.0004$     & 11/16\\
                 & 30 GeV      &--            &$0.123\pm0.006$  & $0.252\pm0.008$  & $2320\pm120$   & $0.019\pm0.004$       & 8/16\\
                 & 40 GeV      & --           &$0.130\pm0.005$  & $0.240\pm0.008$  & $2500\pm119$   & $0.18\pm0.03$         & 12/16\\
                 & 75 GeV     & --            &$0.136\pm0.005$  & $0.240\pm0.009$  & $2700\pm120$   & $2\pm0.3$             & 37/16\\
                 & 150 GeV     & --           &$0.141\pm0.004$  & $0.180\pm0.008$  & $2900\pm130$   & $20\pm4$              & 12/16\\
  \hline
  Fig. 1(d)      & 20 GeV      & $\pi^-$      &$0.133\pm0.005$  & $0.140\pm0.007$  & $2900\pm140$   & $55\pm8$              & 45/12\\
  Pb-Pb          & 30 GeV      & --           &$0.139\pm0.005$  & $0.145\pm0.007$  & $3100\pm152$   &$300\pm32$             & 36/12\\
                 & 40 GeV      &--            &$0.145\pm0.005$  & $0.128\pm0.007$  & $3274\pm147$   & $58\pm9$              & 23/10\\
                 & 80 GeV      & --           &$0.157\pm0.004$  & $0.130\pm0.008$  & $3400\pm138$   & $770\pm100$           & 12/10\\
                 & 160 GeV     & --           &$0.163\pm0.006$  & $0.130\pm0.008$  & $3700\pm143$   & $9000\pm800$          & 15/10\\
  \hline
\end{tabular}%
\end{center}}
\end{table*}

The least square method is used by us in the fit process to get the minimum $\chi^2$. In fig.1, the $\chi^2$ in some cases
are large which show the large dispersion between the curve and data, but the fitting is
approximately acceptable. However in most cases the experimental data is described well by the model in the $p_T$
spectra of the pions produced in different collisions at different energies.

To study the change in the trend of parameters with collision energy and collision cross-section, Fig. 2 shows
the dependencies of kinetic freezeout temperature ($T_0$), transverse flow velocity ($\beta_T$) and kinetic freezeout volume ($V$) on collision energy and collision cross-section. Panel (a), (b) and (c) show the result for kinetic freezeout temperature, transverse flow velocity and kinetic freezeout volume, respectively. Different symbols represent different collision systems. The trend of parameters from left to right shows the energy dependence of the corresponding parameter, while from up to down the different symbols show the collision cross-section dependence of the corresponding parameters. In panel (a), the dependence of kinetic freezeout temperature on the collisions energy and collision cross-section is shown. We observed that as the collision energy increases, $T_0$ also increases. The reason behind it is that at high energies, the collision is very violent which gives a higher excitation to the system. The higher energies, the higher degree of excitation the system will get. Furthermore, it is also observed that $T_0$ in pp collisions is less than in the other three collisions, while in Pb-Pb is the largest, followed by Ar+Sc and then Be-Be collisions. It clearly indicates that $T_0$ depends on the collision cross-section interaction. Larger the collision cross-section, larger $T_0$. Proton-proton is a small collision system and have smaller collision cross-section, while Lead-Lead is the largest collision system and therefore it has  the largest $T_0$. This is in agreement with [14]. Panel (b) shows the dependence of $\beta_T$ on energy and collision cross-section. At the present we observed that $\beta_T$ remains unchanged with increasing the collision energy. We believe that is due to the reason that collective behavior does not change with increasing energy. In addition, $\beta_T$ changes randomly for every system. There is no dependence of $\beta_T$ observed on the collision cross-section. In panel (c), the dependence of $V$ on collision energy and collision cross-section is represented. One can see that $V$ increases as the collision energy increases. The reason behind it is that there is a larger initial bulk system at high energies. The increase in energy results in longer evolution time which corresponds to larger partonic system and the kinetic freezeout volume becomes larger in large partonic system. Furthermore, we also observed that the freezeout volume is larger for Pb-Pb collisions,
followed by Ar-Sc and then Be-Be collisions, and it is the lowest in pp collisions. This clearly indicates its dependence on the collision cross-section (size of the interacting system).

\begin{figure*}
\begin{center}
\includegraphics[width=14.cm]{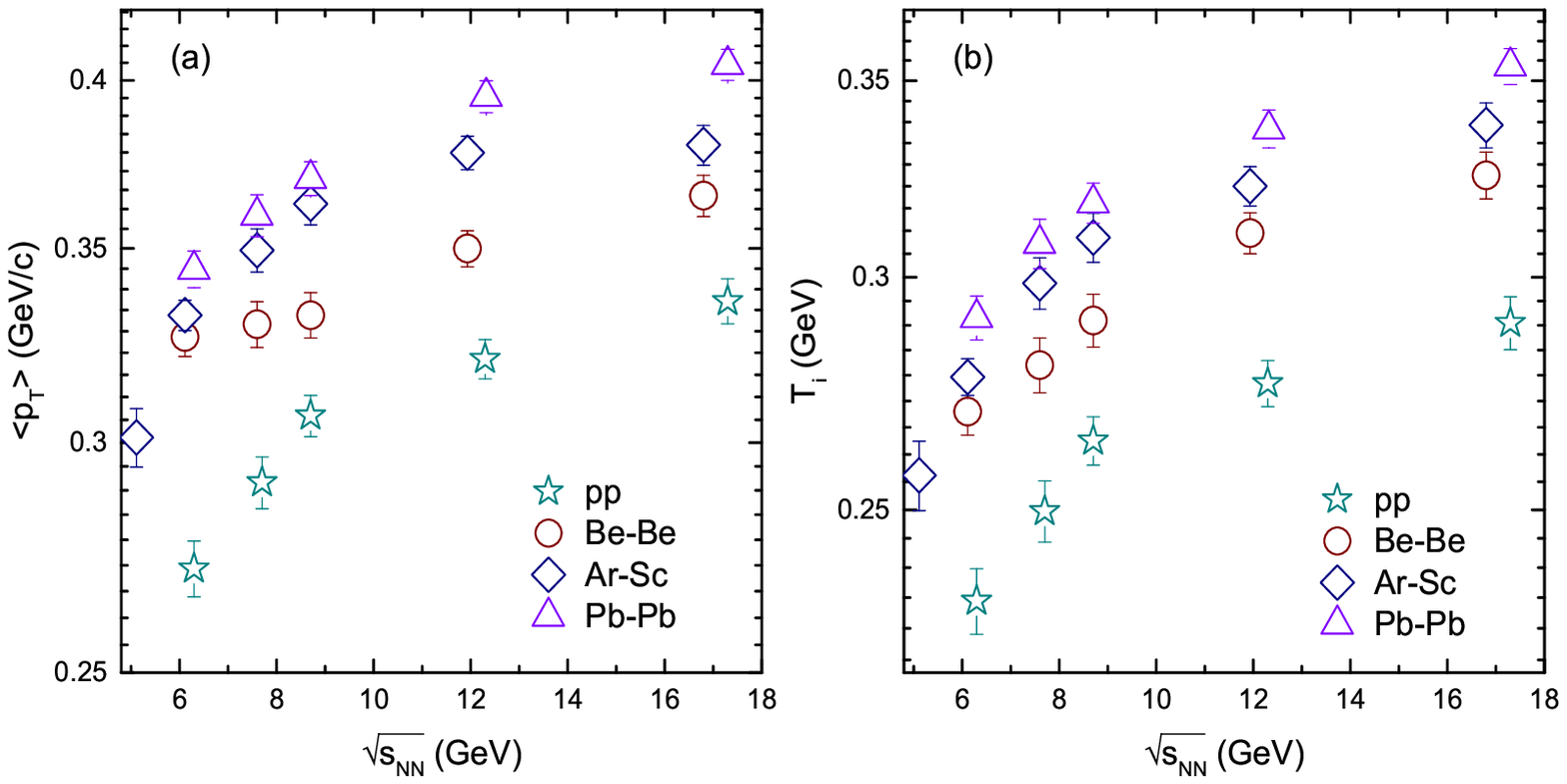}
\end{center}
Fig.3. Dependence of (a) mean transverse momentum, (b) initial temperature on collision energy and collision cross-section.
\end{figure*}

Fig. 3(a) shows the dependence of $<p_T>$ and $T_i$ on collision energy and collision cross-section. One can see that $<p_T>$ increases with an increase of collisions energy and also collision cross-section because larger momentum (energy) is transferred at higher energies and large collision systems, which results in further multiple scattering. However in fig. 3(b), the dependence of initial temperature ($T_i$) on energy and collision cross-section is represented. One can see that $T_i$ increase with increasing the collision energy and is larger for large collision cross-section system.

We observed that the initial temperature is larger than the kinetic freezeout temperature. Generally, the initial temperature is larger than the effective temperature and the effective temperature is larger than the kinetic freezeout temperature because the effective temperature includes the flow effect. The other kind of temperature is chemical freezeout temperature, which is

\begin{align}
T_{ch} = \frac{T_{\lim}}{1+\exp[2.60-\ln(\sqrt{s_{NN}})/0.45]}
\end{align}

where $T_{\lim}=0.1584$ GeV,  and $\sqrt{s_{NN}}$ is in the units of GeV [50]. The chemical freezeout temperature is between the initial and kinetic freezeout temperatures generally, and is slightly larger than or approximately equal to the effective temperature, and this order is in agreement with the order of time evolution of the interacting system.

\section{Summary and Conclusions}

We summarize here our main observations and conclusions as follows.

(a) The transverse momentum spectra of pion produced in inelastic proton-proton, and most central Be-Be, Ar-Sc and Pb-Pb collisions at different energies have been studied by the Blast Wave model with Boltzmann Gibbs statistics. The results are in agreement with the experimental data measured by the NA61/SHINE and NA49
Collaboration at CERN.

(b) Kinetic freeezout temperature and initial temperature increase with the increase of collision energy and collision cross-section
due to the transfer of more energy in the system at higher energies and in large colliding systems.

(c) The transverse flow velocity is observed to remain constant with the increase of energy due to the invariant collective flow with increasing energy. There is no dependence of $\beta_T$ on the collision cross-section.

(d) The kinetic freezeout volume increases with increasing the collision energy due to large initial bulk at higher energies, and it is also larger for large collision systems.

(e) The mean transverse momentum increases with increasing the collision energy due to more transfer of energy in the system at higher energies. It is also observed that  $<p_T>$ is larger for large colliding systems because in large colliding systems the transfer of energy is large.
\\
\\
\\
{\bf Acknowledgments}
This work is supported by the
National Natural Science Foundation of China (Grant
Nos. 11875052, 11575190, and 11135011).
We would also would like to acknowledge the support
of Ajman University Internal Research Grant NO. [DGSR Ref. 2020-IRG-HBS-01].
\\
\\
{\bf Author Contributions} All authors listed have made a
substantial, direct, and intellectual contribution to the work and
approved it for publication.
\\
\\
{\bf Data Availability Statement} This manuscript has no
associated data or the data will not be deposited. [Authors'
comment: The data used to support the findings of this study are
included within the article and are cited at relevant places
within the text as references.]
\\
\\
{\bf Compliance with Ethical Standards}
\\
\\
{\bf Ethical Approval} The authors declare that they are in
compliance with ethical standards regarding the content of this
paper.
\\
\\
{\bf Disclosure} The funding agencies have no role in the design
of the study; in the collection, analysis, or interpretation of
the data; in the writing of the manuscript, or in the decision to
publish the results.
\\
\\
{\bf Conflict of Interest} The authors declare that there are no
conflicts of interest regarding the publication of this paper.
\\
\\

\end{document}